# The degeneracy of the genetic code and Hadamard matrices

Sergey V. Petoukhov

Department of Biomechanics, Mechanical Engineering Research Institute of the
Russian Academy of Sciences
petoukhov@hotmail.com, petoukhov@imash.ru, http://symmetry.hu/isabm/petoukhov.html

**Abstract**: The matrix form of the presentation of the genetic code is described as a cognitive form to analyze structures of the genetic code. A similar matrix form is utilized in the theory of signal processing. The Kronecker family of the genetic matrices is investigated, which is based on the genetic matrix [C A; U G], where C, A, U, G are the letters of the genetic alphabet. This matrix in the third Kronecker power is the (8*8)-matrix, which contains 64 triplets. Peculiarities of the basis scheme of the degeneracy of the genetic code are reflected in the symmetrical black-and-white mosaic of this genetic (8*8)-matrix. This mosaic matrix is connected algorithmically with Hadamard matrices unexpectedly, which are famous in the theory of signal processing, spectral analysis, quantum mechanics and quantum computers. A special decomposition of numeric genetic matrices reveals their close relations with a family of 8-dimensional hypercomplex numbers (not Cayley's octonions). Some hypothesis and thoughts are formulated on the basis of these phenomenological facts.

1. **Introduction**

Genetic information is transferred by means of discrete elements: 4 letters of the genetic alphabet, 64 triplets, 20 amino acids, etc. General theory of signal processing utilizes the encoding of discrete signals by means of special mathematical matrices and spectral representations of signals to increase reliability and efficiency of information transfer [Sklar, 2001; Ahmed, Rao, 1975; etc]. A typical example of such matrices is the family of Hadamard matrices. Rows of Hadamard matrices form an orthogonal system of Walsh functions, which is used for the spectral presentation and transfer of discrete signals [Ahmed, Rao, 1975; Geramita, 1979; Yarlagadda, Hershey, 1997].An investigation of structural analogies between computer informatics and genetic informatics is one of the important tasks of modern science in a connection with a creation of DNA-computers and with a development of bioinformatics. The author investigates molecular structures of the genetic code from the viewpoint of matrix methods of encoding discrete signals.

2. **Method**

This section describes details of the method of the matrix presentation of genetic multiplets to study symmetries and other structural peculiarities of genetic code systems. The utility of the matrix approach for investigations of the genetic code systems was demonstrated in the works [Petoukhov, 2001-2010; He et al., 2004; He, Petoukhov, 2007; Petoukhov, He, 2009].

**Hadamard matrices.** By a definition a Hadamard matrix of dimension "n" is the (n*n)-matrix H(n) with elements "+1" and "-1". It satisfies the condition $H(n)*H(n)^T = n*I_n$, where $H(n)^T$ is the transposed matrix and $I_n$ is the (n*n)-identity matrix. Some of Hadamard matrices of dimension $2^k$ are formed, for example, by the recursive formula $H(2^k) = H(2)^{(k)} = H(2) \otimes H(2^{k-1})$ for $2 \leq k \in N$, where $\otimes$ denotes the Kronecker (or tensor) product, (k) means the Kronecker exponentiation, k and N are integers, H(2) is demonstrated in Figure 1.

Rows of a Hadamard matrix are mutually orthogonal. It means that every two different rows in a Hadamard matrix represent two perpendicular vectors, a scalar product of which is equal to 0. The element "-1" can be disposed in any of four positions in the Hadamard matrix H(2).

$$H(2) = \begin{bmatrix} 1 & 1 \\ -1 & 1 \end{bmatrix} ; \quad H(4) = \begin{bmatrix} 1 & 1 & 1 & 1 \\ -1 & 1 & -1 & 1 \\ -1 & -1 & 1 & 1 \\ 1 & -1 & -1 & 1 \end{bmatrix} ; \quad H(2^k) = \begin{bmatrix} H(2^{k-1}) & H(2^{k-1}) \\ -H(2^{k-1}) & H(2^{k-1}) \end{bmatrix}$$

*Figure 1: The family of Hadamard matrices $H(2^k)$ based on the Kronecker product.*

Such matrices are used in many fields due to their advantageous properties: in error-correcting codes such as the Reed-Muller code; in spectral analysis and multi-channel spectrometers with Hadamard transformations; in quantum computers with Hadamard gates, etc. The author has discovered unexpectedly that Hadamard matrices reflect essential peculiarities of molecular genetic systems.

Normalized Hadamard (2x2)-matrices are matrices of rotation on $45^0$ or $135^0$ depending on an arrangement of signs of its individual elements. A Kronecker product of two Hadamard matrices is a Hadamard matrix as well. A permutation of any columns or rows of a Hadamard matrix leads to a new Hadamard matrix.

Hadamard matrices and their Kronecker powers are used widely in spectral methods of analysis and processing of discrete signals and in quantum computers. A transform of a vector ā by means of a Hadamard matrix H gives the vector ū = H*ā, which is named the Hadamard spectrum of the vector ā usually (but it is possible to call H*ā/4 as the Hadamard spectrum of the vector ā also, and we will use this variant below). A greater analogy between Hadamard transforms and Fourier transforms exists [Ahmed & Rao, 1975]. In particular the fast Hadamard transform exists in parallel with the fast Fourier transform. The whole class of multichannel "spectrometers with Hadamard transforms" is known [Tolmachev, 1976], where the principle of tape masks (or chain masks) is used, and it reminds one of the principles of a chain construction of genetic texts in DNA. Hadamard matrices are used widely in the theory of coding (for example, they are connected with Reed-Muller error correcting codes and with Hadamard codes [Peterson & Weldon, 1972; Solovieva, 2006]; with the theory of compression of signals and images; with a realization of Boolean functions by means of spectral methods; with the theory of planning of multiple-factor experiments and in many other branches of mathematics).

Rows of Hadamard matrices are named Walsh functions or Hadamard functions. Walsh functions can be represented in terms of product of Rademacher functions $r_n(t) = sign(sin2^n\pi t)$, n = 1,2,3,…, which accept the two values "+1" and "-1" only (here "sign" is the function of a sign on argument) [Ahmed, Rao, 1975]. Sets of numerated Walsh functions, when they are united in square matrices, form systems depending on features of such union. Figure 2 shows two examples of systems of such functions, which are used widely in the theory of digital signals processing.

They are connected with Hadamard (8x8)-matrices and with the Walsh-Hadamard transform, which is the most famous among non-sinusoidal orthogonal transforms and which can be calculated by means of mathematical operations of addition and subtraction only (see more detail in [Ahmed & Rao, 1975; Trahtman & Trahtman, 1975; Yarlagadda, & Hershey, 1997]. Hereinafter we will use the simplified designations of matrix elements on illustrations of Hadamard matrices: the symbol "+" or the black color of a matrix cell means the element "+1"; the symbol "-" or the white color of a matrix cell means the element "-1". The theory of discrete signals pays a special attention to quantities of changes of signs "+" and "-" along each row and each column in Hadamard matrices. These quantities are connected with an important notion of "sequency" as a generalization of notion of "frequency" [Ahmed & Rao, 1975, p.85; Harmut, 1989]. Figure 2 shows these quantities for each row and each column in presented matrix examples. Each of these two Hadamard matrices is symmetrical relative to its main diagonal.

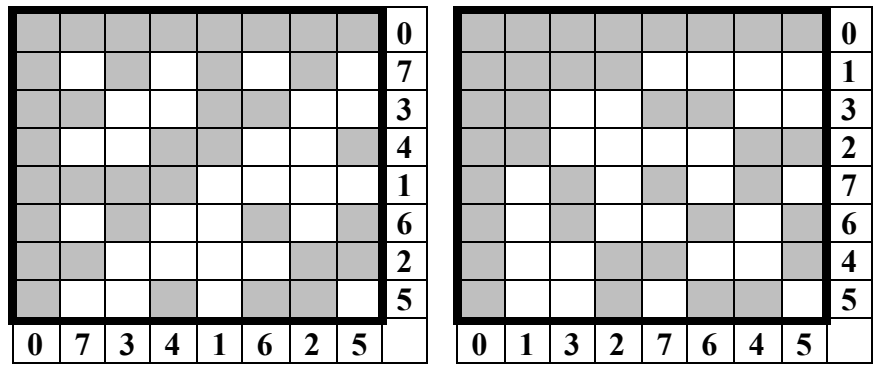

*Figure 2. Examples of two systems of Walsh functions, which are used frequently in the theory of digital signal processing [Trahtman & Trahtman, 1975]. On the left side: the Walsh-Hadamard system. On the right side: the Walsh-Paley system. Quantities of changes of signs "+" and "-" are shown for each row and each column.*

Some of normalized Hadamard matrices are unitary operators (for example $2^{-0.5}*[1\ 1;\ 1\ -1]$ and $2^{-0.5}*[1\ -1;\ 1\ 1]$). They serve as one of the important instruments to create quantum computers, which utilize so called Hadamard gates (as evolution of the closed quantum system is unitary) [Nielsen & Chuang, 2001]. Let us demonstrate now unexpected connections of Hadamard matrices with Kronecker families of genetic matrices.

**The Kronecker family of genetic matrices.** Genetic multiplets are one of main peculiarities of the genetic code. Really, the alphabet of the genetic code is a set of 4 monoplets (nitrogenous bases): A (adenine), C (cytosine), G (guanine), U/T (uracil in RNA or thymine in DNA); 64 triplets encode amino acids; each protein is encoded by more or less long multiplets. On the base of the idea about analogies between computer informatics and genetic informatics, the author has represented all sets of genetic multiplets as proper parts of the Kronecker family of the square matrices (genomatrices) $P^{(n)} = [C\ A;\ U\ G]^{(n)}$, where A, C, G, U are the letters of the genetic alphabet, (n) means the Kronecker exponentiation.

Each genomatrix [C A; U G]$^{(n)}$ contains a complete set of n-plets as its matrix elements. For example, the (8x8)-genomatrix [C A; U G]$^{(3)}$ contains all 64 triplets which encode 20 amino acids and stop-signals. This family of matrices gives complete sets of n-plets in the universal mathematical form, which is based on the square matrix of the genetic alphabet.

$$P = \begin{bmatrix} C & A \\ U & G \end{bmatrix} ; \quad P^{(2)} = P \otimes P = \begin{bmatrix} CC & CA & AC & AA \\ CU & CG & AU & AG \\ UC & UA & GC & GA \\ UU & UG & GU & GG \end{bmatrix}$$

$$P^{(3)} = \begin{bmatrix} CCC & CCA & CAC & CAA & ACC & ACA & AAC & AAA \\ CCU & CCG & CAU & CAG & ACU & ACG & AAU & AAG \\ CUC & CUA & CGC & CGA & AUC & AUA & AGC & AGA \\ CUU & CUG & CGU & CGG & AUU & AUG & AGU & AGG \\ UCC & UCA & UAC & UAA & GCC & GCA & GAC & GAA \\ UCU & UCG & UAU & UAG & GCU & GCG & GAU & GAG \\ UUC & UUA & UGC & UGA & GUC & GUA & GGC & GGA \\ UUU & UUG & UGU & UGG & GUU & GUG & GGU & GGG \end{bmatrix}$$

*Figure 3: The beginning of the Kronecker family of symbolic genomatrices $P^{(n)} = [C A; U G]^{(n)}$ for n = 1, 2, 3.*

**The black-and-white matrix mosaic of the basic scheme of the degeneracy of the genetic code.** The genetic code is named "the degeneracy code" because its 64 triplets encode 20 amino acids and different amino acids are encoded by different quantities of triplets. Hypotheses about a connection between this degeneracy and the noise-immunity of genetic information exist since the time of the discovery of the genetic code. The specifics of the degeneracy of the genetic code provoke many questions. One of them is the following: was the code degeneracy an accidental choice of nature or not? Deep investigations of symmetries using a matrix map of the code degeneracy can give many useful results for such questions.

Modern science recognizes many variants (or dialects) of the genetic code, data about which are shown on the NCBI's website http://www.ncbi.nlm.nih.gov/Taxonomy/Utils/wprintgc.cgi. 17 variants (or dialects) of the genetic code exist which differ one from another by some details of correspondences between triplets and objects encoded by them. Most of these dialects (including the so called Standard Code and the Vertebrate Mitochondrial Code) have the following general scheme of their degeneracy where 32 "black" triplets with "strong roots" and 32 "white" triplets with "weak" roots exist (the terms "strong roots" and "weal roots" of triplets were introduced by Yu. Rumer (1968)).

In this general or basic scheme, the set of 64 triplets contains 16 subfamilies of triplets, every one of which contains 4 triplets with the same two letters on the first positions of each triplet (an example of such subsets is the case of the four triplets CAC, CAA, CAU, CAG with the same two letters CA on their first positions). We shall name such subfamilies as the subfamilies of NN-triplets. In the described basic scheme of degeneracy, the set of these 16 subfamilies of NN-triplets is divided into two equal subsets from the viewpoint of degeneration properties of the code. The first subset contains 8 subfamilies of so called "two-position" NN-triplets, a coding value of which is independent of a letter on their third position (Figure 4 illustrates the coding meanings of the subfamilies of NN-triplets for the case of the Vertebrate Mitochondrial Code). An example of such subfamilies is the four triplets CGC, CGA, CGU, CGC, all of which encode the same amino acid Arg, though they have different letters on their third position. All members of such subfamilies of NN-triplets are marked by black color in Figure 4.

The second subset contains 8 subfamilies of "three-position" *NN*-triplets, the coding value of which depends on a letter on their third position. An example of such subfamilies in Figure 4 is the four triplets CAC, CAA, CAU, CAC, two of which (CAC, CAU) encode the amino acid His and the other two (CAA, CAG) encode another amino acid Gln. All members of such subfamilies of *NN*-triplets are marked by the white color in the genomatrix $P^{(3)} = [C\ A;\ U\ G]^{(3)}$ on Figure 4. So the genomatrix $[C\ A;\ U\ G]^{(3)}$ has 32 black triplets and 32 white triplets. Each subfamily of four *NN*-triplets is disposed in an appropriate (2x2)-subquadrant of the genomatrix $[C\ A;\ U\ G]^{(3)}$ due to the Kronecker algorithm of construction of the (8x8)-genomatrix $[C\ A;\ U\ G]^{(3)}$ of triplets from the alphabet (2x2)-genomatrix $[C\ A;\ U\ G]$ (Figure 3).

One can check easily on the basis of data from the NCBI's website (http://www.ncbi.nlm.nih.gov/Taxonomy/Utils/wprintgc.cgi) that the following 11 dialects of the genetic code have the same basic scheme of degeneracy with 32 black triplets and with 32 white triplets: 1) the Standard Code; 2) the Vertebrate Mitochondrial Code; 3) the Yeast Mitochondrial Code; 4) the Mold, Protozoan, and Coelenterate Mitochondrial Code and the Mycoplasma/Spiroplasma Code; 5) the Ciliate, Dasycladacean and Hexamita Nuclear Code; 6) the Euplotid Nuclear Code; 7) the Bacterial and Plant Plastid Code; 8) the Ascidian Mitochondrial Code; 9) the Blepharisma Nuclear Code; 10) the Thraustochytrium Mitochondrial Code; 11) the Chlorophycean Mitochondrial Code. In this article we will consider this basic scheme of the degeneracy which is presented by means of a black-and-white mosaic of the genetic matrix $[C\ A;\ U\ G]^{(3)}$ on Figure 4.

One can mentioned that the other 6 dialects of the genetic code have only small differences from the described basic scheme of degeneracy: the Invertebrate Mitochondrial Code; the Echinoderm and Flatworm Mitochondrial Code; the Alternative Yeast Nuclear Code; The Alternative Flatworm Mitochondrial Code; the Trematode Mitochondrial Code; the Scenedesmus obliquus mitochondrial Code.

The disposition of these "black" and "white" triplets forms the very symmetric black-and-white mosaic in the genomatrix $[C\ A;\ U\ G]^{(3)}$ (Figure 4). For instance, left and right matrix halves are mirror-anti-symmetric to each other in its colors: any pair of cells, disposed by the mirror-symmetrical manner in these halves, has opposite colors. Diagonal quadrants of the matrix are identical to each other from the viewpoint of their mosaic. The rows 1-2, 3-4, 5-6, 7-8 are identical to each other from the viewpoint of the mosaic, etc. A sequence of black and white cells in each row has a meander character and corresponds to one of Rademacher functions (Figure 4, on the right side). One can say that the mosaic of this genetic matrix has a natural numeric presentation in a form of a set of Rademacher functions.

It should be mentioned that the quantity of variants of possible dispositions of 64 genetic triplets in 64 cells of a (8*8)-matrix is equal to the huge number $64! \sim 10^{89}$. It is obvious that the most of these variants have not symmetries between mosaics of matrix quadrants and have not a close relation with Rademacher functions.

| CCC | CCA | CAC | CAA | ACC | ACA | AAC | AAA |
| PRO | PRO | HIS | GLN | THR | THR | ASN | LYS |
| CCU | CCG | CAU | CAG | ACU | ACG | AAU | AAG |
| PRO | PRO | HIS | GLN | THR | THR | ASN | LYS |
| CUC | CUA | CGC | CGA | AUC | AUA | AGC | AGA |
| LEU | LEU | ARG | ARG | ILE | MET | SER | STOP |
| CUU | CUG | CGU | CGG | AUU | AUG | AGU | AGG |
| LEU | LEU | ARG | ARG | ILE | MET | SER | STOP |
| UCC | UCA | UAC | UAA | GCC | GCA | GAC | GAA |
| SER | SER | TYR | STOP | ALA | ALA | ASP | GLU |
| UCU | UCG | UAU | UAG | GCU | GCG | GAU | GAG |
| SER | SER | TYR | STOP | ALA | ALA | ASP | GLU |
| UUC | UUA | UGC | UGA | GUC | GUA | GGC | GGA |
| PHE | LEU | CYS | TRP | VAL | VAL | GLY | GLY |
| UUU | UUG | UGU | UGG | GUU | GUG | GGU | GGG |
| PHE | LEU | CYS | TRP | VAL | VAL | GLY | GLY |

*Figure 4: The representation of the genomatrix [C A; U G]$^{(3)}$ of 64 triplets for the case of the vertebrate mitochondrial genetic code. The matrix contains 20 amino acids with their traditional abbreviations. Stop-codons are marked as "stop". Relevant meander functions by Rademacher are shown for each matrix row on the right side. Matrix cells with "black" ("white") triplets are marked by dark (white) color.*

Why the nature has chosen (from the huge number of possible variants) this specific code degeneracy, which is reflected in such symmetric mosaic and in such relation with Rademacher functions? In author's opinion, which will be supported below, a key meaning for receiving an effective answer on this question belongs to a fact of the connection of this mosaic genetic matrix with Rademacher functions which are related with Walsh functions from Hadamard matrices. Taking into account this known relation between Rademacher and Walsh functions, one can think about a hidden connection between the mosaic of the genetic matrix [C A; U G]$^{(3)}$ and an orthogonal system of Walsh functions from Hadamard matrices. Does a simple algorithmic connection exists between the genetic matrix [C A; U G]$^{(3)}$ and a Hadamard (8*8)-matrix?

| +1 | +1 | -1 | -1 | +1 | +1 | -1 | -1 |
| +1 | +1 | -1 | -1 | +1 | +1 | -1 | -1 |
| +1 | +1 | +1 | +1 | -1 | -1 | -1 | -1 |
| +1 | +1 | +1 | +1 | -1 | -1 | -1 | -1 |
| +1 | +1 | -1 | -1 | +1 | +1 | -1 | -1 |
| +1 | +1 | -1 | -1 | +1 | +1 | -1 | -1 |
| -1 | -1 | -1 | -1 | +1 | +1 | +1 | +1 |
| -1 | -1 | -1 | -1 | +1 | +1 | +1 | +1 |

→

| +1 | +1 | -1 | -1 | +1 | +1 | -1 | -1 | *3* |
| -1 | +1 | +1 | -1 | -1 | +1 | +1 | -1 | *4* |
| +1 | +1 | +1 | +1 | -1 | -1 | -1 | -1 | *1* |
| -1 | +1 | -1 | +1 | +1 | -1 | +1 | -1 | *6* |
| -1 | -1 | +1 | +1 | +1 | +1 | -1 | -1 | *2* |
| +1 | -1 | -1 | +1 | -1 | +1 | +1 | -1 | *5* |
| +1 | +1 | +1 | +1 | +1 | +1 | +1 | +1 | *0* |
| -1 | +1 | -1 | +1 | -1 | +1 | -1 | +1 | *7* |
| *5* | *2* | *6* | *1* | *5* | *2* | *6* | *1* | |

*Figure 5. Top: the genomatrix [C A; U G]$^{(3)}$ in its numeric Rademacher form R of presentation. Bottom: the Hadamard matrix $H_u$ which is received from [C A; U G]$^{(3)}$ by means of the U-algorithm (a relevant orthogonal system of Walsh functions are shown on the right side).*

*Numbers of sequences (quantities of changes of signs of matrix components) for each row and for each column are shown additionally.*

Below we shall demonstrate an existence of such algorithm based on molecular peculiarities of the genetic alphabet. For this demonstration we will consider the following numeric presentation R ("Rademacher presentation") of the mosaic genomatrix [C A; U G]$^{(3)}$ which reflects its relation with Rademacher functions: we will place the signs "+1" and "-1" into black and white cells of the genomatrix on Figure 4 correspondingly. In such presentation we receive the numeric matrix G on Figure 5 (top). The genomatrix [C A; U G]$^{(3)}$ is transformed into the Hadamard matrix $H_u$ on Figure 5 (bottom) by the simple algorithm described in the next section.

**Amino-groups NH$_2$ and the U-algorithm.** It is known that amino-groups NH$_2$ play an important role in molecular genetics. For instance, the amino-group in amino acids provides a function of recognition of the amino acids by ferments [Chapeville, Haenni, 1974]. A detachment of amino-groups in nitrogenous bases A and C in RNA under action of nitrous acid HNO$_2$ determines a property of amino-mutating of these bases, which was used to divide the set of 64 triplets in eight natural subsets with 8 triplets in each [Wittmann, 1961].

But how the amino-groups are represented in the genetic alphabet? One can note that each of three nitrogenous bases A, C, G has one amino-group, but the fourth basis U/T has not it (Figure 6). From the viewpoint of existence of the amino-group, the letters A, C, G are identical to each other and the letter U is opposite to them.

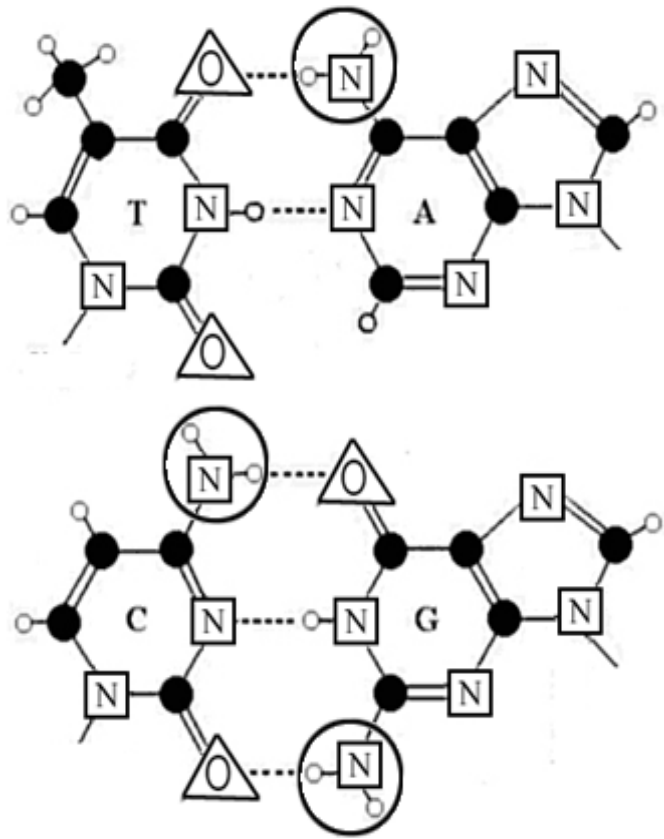

*Figure 6. The complementary pairs of the four nitrogenous bases in DNA. A-T (adenine and thymine), C-G (cytosine and guanine). Hydrogen bonds in these pairs are shown by dotted lines. Black circles are atoms of carbon; small white circles are atoms of hydrogen; squares with the letter N are atoms of nitrogen; triangles with the letter O are atoms of oxygen. Amides (or amino-groups) NH$_2$ are marked by big circles.*

This fact of existence or absence of the amino-group in certain genetic letters can be reflected in the alphabetic genomatrix P=[C A; U G] by symbols "+1" and "-1" instead of the letters A, C, G and U/T correspondingly. In this case the Hadamard genomatrix $P_{H(2)}$ = H(2) appears (Figure 7).

$$P = \begin{bmatrix} C & A \\ U & G \end{bmatrix} \rightarrow H(2) = \begin{bmatrix} +1 & +1 \\ -1 & +1 \end{bmatrix}$$

*Figure 7: The transformation of the matrix P=[C A; U G] of the genetic alphabet into the Hadamard matrix. Black cells of this Hadamard matrix contain elements "+1", and the white cell contains the element "-1".*

Hence the letter U in RNA (and the letter T in DNA) is the peculiar letter in the genetic alphabet in this sense. The letter U has also another unique property among 4 letters of the genetic alphabet: the letter U exists in RNA, but in DNA it is replaced by the letter T. These molecular properties of the letter U can be utilized in genetic computers of organisms. Taking into account this unique status of the letter U, the author revealed the existence of the following formal "U-algorithm", which demonstrates the close connection between Hadamard matrices and the matrix mosaic of the degeneracy of the genetic code.

The definition of the U-algorithm: each triplet in the black-and-white genomatrix [C A; U G]$^{(3)}$ on Figure 4 should change its own color into opposite color each time when the letter U stands in an odd position (in the first or in the third position) inside the triplet. For example, the white triplet UUA (see Figure 4) should become the black triplet (and its matrix cell should be marked by black color) because of the letter U in its first position. Or the white triplet UUU should not change its color because of the letter U in its first and third positions (the color of this triplet is changed twice according to the described algorithm). The triplet ACG does not change its color because the letter U is absent in this triplet at all. By means of the U-algorithm, the genomatrix $P^{(3)}$ = [C A; U G]$^{(3)}$ from Figure 4 is transformed into the genomatrix $P_H^{(3)}$ on Figure 7. At the last step of the U-algorithm all triplets in black (white) cells are replaced by elements "+1" ("-1") to receive the Hadamard matrix $H_u$ from Figure 5.

| CCC | CCA | CAC | CAA | ACC | ACA | AAC | AAA |
|-----|-----|-----|-----|-----|-----|-----|-----|
| CCU | CCG | CAU | CAG | ACU | ACG | AAU | AAG |
| CUC | CUA | CGC | CGA | AUC | AUA | AGC | AGA |
| CUU | CUG | CGU | CGG | AUU | AUG | AGU | AGG |
| UCC | UCA | UAC | UAA | GCC | GCA | GAC | GAA |
| UCU | UCG | UAU | UAG | GCU | GCG | GAU | GAG |
| UUC | UUA | UGC | UGA | GUC | GUA | GGC | GGA |
| UUU | UUG | UGU | UGG | GUU | GUG | GGU | GGG |

*Figure 8: The mosaic genomatrix $P_H^{(3)}$, which is received from [C A; U G]$^{(3)}$ by the U-algorithm.*

## 3. Results

The genomatrix $P_H^{(3)}$ (Figure 8) possesses the black-and-white mosaic which is identical to the mosaic of the matrix $H_u$ on Figure 5. The matrix $H_u$ is one of Hadamard matrices because it satisfies the general condition of Hadamard matrices $H(n)*H(n)^T = n*I_n$.

The Hadamard matrix $H_u$ has the interesting property of a fractal character. Each (4*4)-quadrant of the (8*8)-matrix $H_u$ is a Hadamard matrix also. Furthermore, each (2*2)-sub-quadrant of the

(8*8)-matrix $H_u$ is a Hadamard matrix as well (this situation can be named "Hadamard fractals" conditionally).

| CCC Pro | CAC His | ACC Thr | AAC Asn | CCA Pro | CAA Gln | ACA Thr | AAA Lys |
|---|---|---|---|---|---|---|---|
| CUC Leu | CGC Arg | AUC Ile | AGC Ser | CUA Leu | CGA Arg | AUA Met | AGA Stop |
| UCC Ser | UAC Tyr | GCC Ala | GAC Asp | UCA Ser | UAA Stop | GCA Ala | GAA Glu |
| UUC Phe | UGC Cys | GUC Val | GGC Gly | UUA Leu | UGA Trp | GUA Val | GGA Gly |
| CCU Pro | CAU His | ACU Thr | AAU Asn | CCG Pro | CAG Gln | ACG Thr | AAG Lys |
| CUU Leu | CGU Arg | AUU Ile | AGU Ser | CUG Leu | CGG Arg | AUG Met | AGG Stop |
| UCU Ser | UAU Tyr | GCU Ala | GAU Asp | UCG Ser | UAG Stop | GCG Ala | GAG Glu |
| UUU Phe | UGU Cys | GUU Val | GGU Gly | UUG Leu | UGG Trp | GUG Val | GGG Gly |

Figure 9: The genomatrix $P_{231}^{(3)}$.

The theory of signal processing pays a special attention to permutations of components. Let us analyze properties of the genomatrix $P^{(3)} = [C\ A;\ U\ G]^{(3)}$ relative to positional permutations in all triplets. Any triplet has the six permutation variants of a sequence of its three positions: 1-2-3, 2-3-1, 3-1-2, 1-3-2, 2-1-3, 3-2-1. Let us take for example the following permutation of the cyclic shift: 1-2-3 → 2-3-1. In the result of this permutation the triplet CAG is replaced in its matrix cell by the triplet AGC, etc. And the whole genomatrix $P_{123}^{(3)} = [C\ A;\ U\ G]^{(3)}$ is reconstructed cardinally into the new mosaic matrix $P_{231}^{(3)}$ (Figure 9), which has considerable symmetries also (all quadrants have the identical mosaics; upper and bottom halves have identical contents, etc). Mosaics of all rows of this new matrix $P_{231}^{(3)}$ correspond to a set of Rademacher functions again.

The other four variants of the positional permutations in the triplets produce the four genomatrices $P_{312}^{(3)}$, $P_{132}^{(3)}$, $P_{213}^{(3)}$, $P_{321}^{(3)}$ (the bottom indexes show the positional permutations), each of which has considerable symmetries also. Mosaics of all rows of these new genomatrices correspond to Rademacher functions again. It means that the degeneracy of the genetic code is connected closely with Rademacher functions and with the positional permutations in the set of 64 triplets. The same U-algorithm transforms the mosaics of the matrices $P_{231}^{(3)}$, $P_{312}^{(3)}$, $P_{132}^{(3)}$, $P_{213}^{(3)}$, $P_{321}^{(3)}$ into new mosaics, which coincide with the mosaics of the proper Hadamard matrices on Figure 10. One can note additionally a special feature connected with quantities of changes of the signs "+" and "-" in rows and columns of all genetic Hadamard matrices on Figures 5 and 10. Each sum of such quantities in the first four rows and in second four rows of such matrix is equal to 14. And each sum of such quantities in the first four columns and in second four columns of such matrix is equal to 14. It can be named a phenomenon of a balanced sequency in these genetic Hadamard matrices (the notion of "sequency" is a generalization of notion of "frequency" [Ahmed & Rao, 1975, p.85]). It should be noted that Hadamard matrices, which usually are used in the theory of signal processing (see Figure 2), do not possess such feature. In addition all genetic Hadamard matrices on Figures 5 and 10 are asymmetric relative to both diagonals in contrast to the technical Hadamard matrices on Figure 2.

Besides the described cyclic permutations of positions in the triplets (123→231→312 and 321→213→132), a few other types of cyclic permutations of the genetic elements exist, which reveal new genetic Hadamard matrices. The speech is about the fact that each of such permutations transforms the initial genetic matrix [C A; U G]$^{(3)}$ into a new genetic mosaic matrix which, firstly, is connected with Rademacher functions by analogy with above examples and which, secondly, can be transformed into a relevant Hadamard matrix by the same U-algorithm. For example, one of such types of permutations is a cyclic alphabetical permutation of the genetic letters C→G→U→A→C. The application of this alphabetical permutation to the initial genomatrix $P^{CAUG}_{123}$ = [C A; U G]$^{(3)}$ transforms it into a new genomatrix $P^{GCAU}_{123}$ = [G C; A U]$^{(3)}$. The permutations of positions in the triplets in this genomatrix $P^{GCAU}_{123}$ leads to new genomatrices $P^{GCAU}_{231}$, $P^{GCAU}_{312}$, $P^{GCAU}_{321}$, $P^{GCAU}_{213}$, $P^{GCAU}_{132}$ by analogy with the case considered above (see Figure 10). The same U-algorithm transforms all these genomatrices into genetic Hadamard matrices presented on Figure 11. The second part of Figure 11 demonstrates additionally a case of Hadamard matrices which are received by means of the U-algorithm from genomatrices $P^{CAGU}_{123}$ = [C A; G U]$^{(3)}$, $P^{CAGU}_{231}$, $P^{CAGU}_{312}$, $P^{CAGU}_{321}$, $P^{CAGU}_{213}$ and $P^{CAGU}_{132}$.

All kinds of genetic Hadamard matrices, which were met by the author in this study, possess the following analogical features with the Hadamard matrices on Figures 5 and 10. Firstly, each of such genetic Hadamard matrices is asymmetrical relative to both diagonals. Secondly, each of them demonstrates the described phenomenon of a balanced sequency as well. More details on this theme can be found in the books [Petoukhov, 2008; Petoukhov, He, 2009].

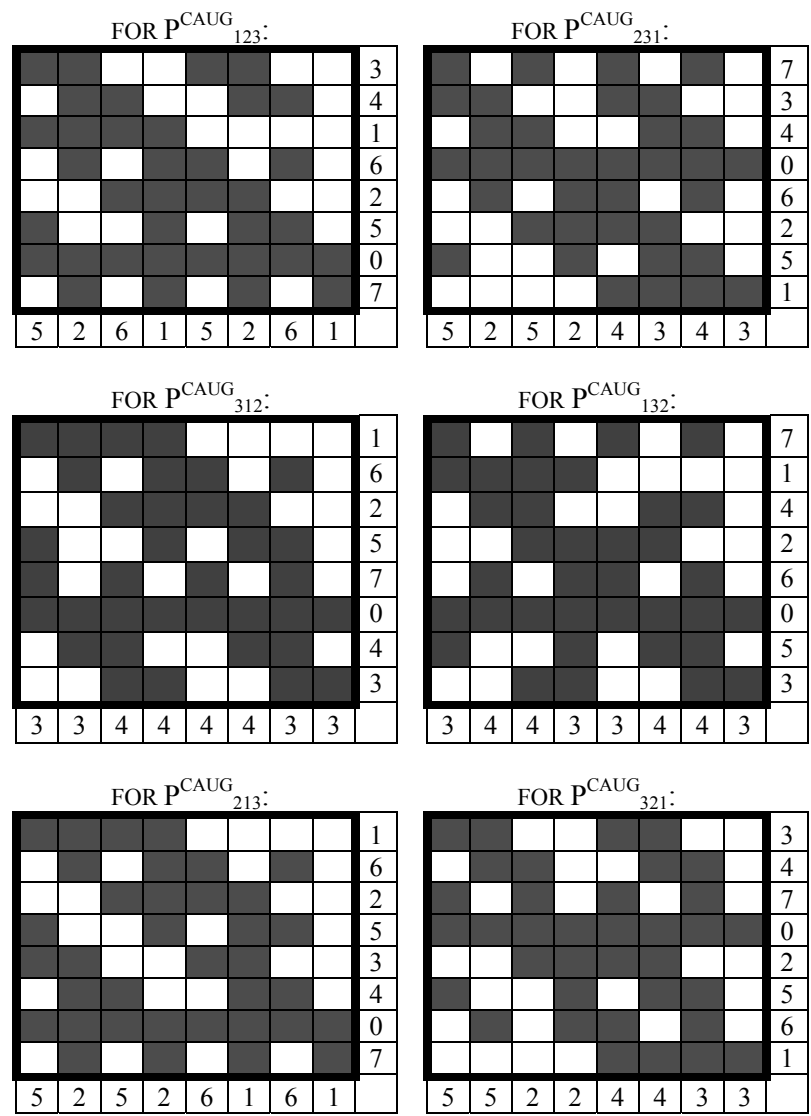

*Figure 10. The six Hadamard matrices, which are produced from the six indicated genomatrices by means of the U-algorithm. The black cells correspond to elements "+1" and the white cells correspond to elements "-1". Quantities of changes of signs "+" and "-" (or changes of colors) are shown for each row and each column.*

Infinite number of systems of orthogonal functions exists which can be used to create different kinds of spectral analysis. Why we should pay a special attention to Hadamard matrices and their Walsh functions? The author thinks that described genetic Hadamard matrices and relevant systems of Walsh functions are important objects for molecular genetics and theoretical biology by the following main reasons:

- Known advantages of use of Walsh functions in digital communication are connected, firstly, with a fact that a computer realization of these functions, which consists of two elements "+1" and "-1" only, is simpler than a realization of trigonometric functions for Fourier analysis and, secondly, with a fact that their applications for spectral analysis can be based on mathematical operations of addition and subtraction only. It seems that the operations of addition and subtraction can be realized inside bio-molecular computers simply by means of a creation of new chemical bonds and by means of a destruction of chemical bonds. In other words, these operations are more simple and natural for ensembles of biological molecules in comparison with operations of multiplication and division. Consequently applications of Walsh functions fit to a set of genetic molecules more than applications of trigonometric functions, etc.

- Biological molecules are subordinated to laws of quantum mechanics. One can think that spectral methods are essential for electro-magnetic and other kinds of communications among these molecules. By this reason we should pay a special attention to those systems of orthogonal functions which have natural connections with operators of quantum mechanics. Normalized Hadamard operators belong to such operators because some of them are unitary operators of quantum mechanics (evolution of a closed quantum system is unitary). One can remember here that quantum computers are created on a basis of Hadamard gates not without reason. So one should study a possible meaning of Hadamard matrices for molecular genetics intensively. But many kinds of Hadamard matrices exist. For example a quantity of Hadamard (8*8)-matrices is equal to $5*10^9$ (five billion) approximately. What kinds of Hadamard matrices should be studied in the genetic field preferably? Matrix genetics reveals an existence of special families of Hadamard matrices which are connected with genetic matrices. These families should be investigated first of all.

It is known an application of Walsh functions in molecular genetics as one of mathematical tools for investigation of cyclical recurrences in genetic sequences [Tavare, Giddings, 1989]. But this application does not exhaust possible meanings of Hadamard matrices and Walsh functions in the field of molecular genetics. Below we will speak about one of possible variants of new applications of Hadamard matrices in molecular genetics and bioinformatics for new tasks.

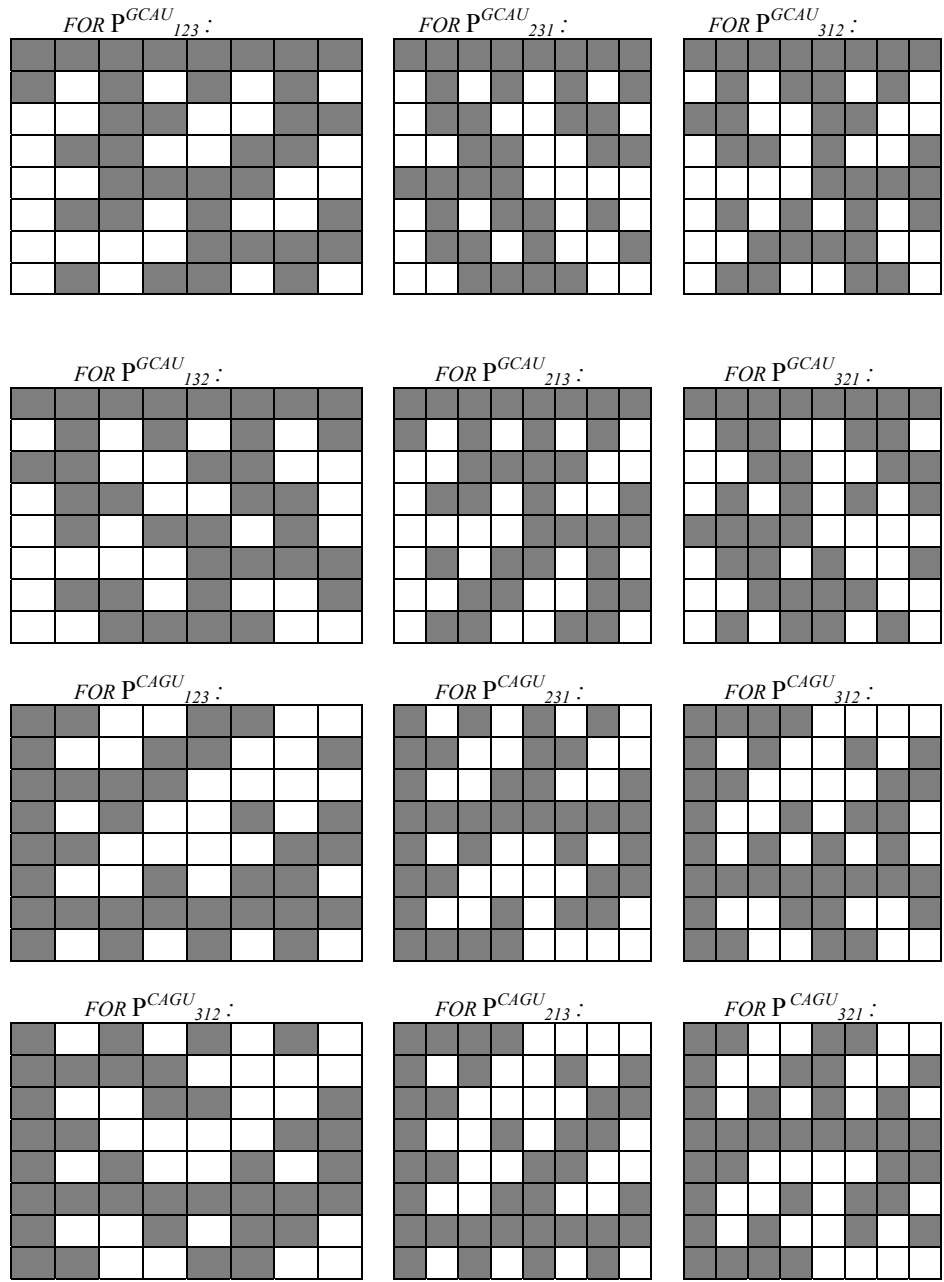

*Figure 11. The 12 balanced Hadamard matrices, which are produced from the indicated 12 genomatrices of triplets by means of the U-algorithm. Black cells correspond to elements "+1" and white cells correspond to elements "-1".*

## 4. Bi-sign vectors and Hadamard matrices.

Let us consider now a set of 8-dimensional vectors, each component of which is equal to "+1" or "-1". For example this set contains vectors [+1; +1; -1; +1; -1; -1; +1;+1], [-1;-1; +1;+1;+1;+1;+1;+1], etc. We will call such vector as "bi-sign 8-dimensional vector" or "a spin-like 8-dimensional vector" (in a general case a notion of bi-sign $2^k$-dimensional vectors is possible and useful). Quantity of elements "+1" in a bi-sign vector is called its "weight". For example, the weight of the vector [+1; +1; -1; +1; -1; -1; +1;+1] is equal to 5 and the weight of the vector [-1; -1; -1; -1; -1; -1; -1;-1] is equal to 0. The set of the bi-sign 8-dimensional vectors is divided into two sub-sets: one of which contains vectors with even weights (these weights are equal to even numbers 0, 2, 4, 6, 8) and the second of which contains vectors with odd weights (these weights are equal to odd numbers 1, 3, 5, 7). By this reason, bi-sign vectors of the first type are called "even bi-sign vectors" and bi-sign vectors of the second type are called "odd bi-sign vectors".

An interesting property of Hadamard (8*8)-matrices is connected with its multiplication with bi-sign 8-dimensional vectors. For example let us consider a result of multiplication $H_u^T*\bar{e}/2$ of the transposed genetic Hadamard matrix $H_u^T$ (the matrix $H_u$ is from Fig. 5) with an arbitrary vector $\bar{e}$ = [$a_0$; $a_1$; $a_2$; $a_3$; $a_4$; $a_5$; $a_6$; $a_7$]:

$$H_u^T*\bar{e}/2 = 0,5*[a_0-a_1+a_2-a_3-a_4+a_5+a_6-a_7$$
$$a_0+a_1+a_2+a_3-a_4-a_5+a_6+a_7$$
$$-a_0+a_1+a_2-a_3+a_4-a_5+a_6-a_7$$
$$-a_0-a_1+a_2+a_3+a_4+a_5+a_6+a_7$$
$$a_0-a_1-a_2+a_3+a_4-a_5+a_6-a_7$$
$$a_0+a_1-a_2-a_3+a_4+a_5+a_6+a_7$$
$$-a_0+a_1-a_2+a_3-a_4+a_5+a_6-a_7$$
$$-a_0-a_1-a_2-a_3-a_4-a_5+a_6+a_7] \quad (1)$$

If each coordinate $a_i$ is equal to +1 or -1 ($a_0 = a_1=\ldots= a_7=\pm 1$), then one can check easily from the expression (1) that the whole set of final vectors $H_u^T*\bar{e}_i/2$ is divided into two opposite sub-sets by the dichotomy principle "even-odd". One of these sub-sets is based on the odd type of bi-sign vectors $\bar{e}_i$, and its final vectors $H_u^T*\bar{e}_i/2$ have their coordinates only with odd numbers $\pm 1$ and $\pm 3$. By this reason the final vectors of the first sub-set can be called as "odd vectors $H_u^T*\bar{e}_i/2$". The second of these subsets is based on the even type of bi-sign vectors $\bar{e}_i$, and its final vectors have coordinates only with even numbers 0, $\pm 2$ and $\pm 4$. By this reason the final vectors of the second subset can be called as "even vectors $H_u^T*\bar{e}_i/2$". Figure 12 shows some examples of such operator action of the genetic Hadamard matrix $H_u^T$ with the factor 1/2 on some 8-dimensional bi-signs vectors. The initial bi-sign vectors $\bar{e}_i$ can be reconstructed easily from the vectors $H_u^T*\bar{e}_i/2$ by means of the operation $H_u*(H_u^T*\bar{e}_i/2)/4$ (in accordance with methods of spectral analysis on the basis of Hadamard matrices).

| A case of bi-sign vectors $\bar{e}_i$ with even weights | A case of bi-sign vectors $\bar{e}_i$ with odd weights |
|---|---|
| $H_u^T*[1;1;1;1;1;1;1;1]/2 =[0; 2; 0; 2; 0; 2; 0;-2]$ | $H_u^T*[-1;1;1;1;1;1;1;1]/2=[-1; 1; 1; 3;-1; 1; 1;-1]$ |
| $H_u^T*[-1;-1;1;1;1;1;1;1]/2 =[0; 0; 0; 4; 0; 0; 0; 0]$ | $H_u^T*[-1;-1;-1;1;1;1;1;1]/2 =[-1;-1;-1; 3; 1; 1; 1; 1]$ |
| $H_u^T*[-1;-1;-1;-1;1;1;1;1]/2 =[0;-2; 0; 2; 0; 2; 0; 2]$ | $H_u^T*[-1;-1;-1;-1;-1;1;1;1]/2=[1;-1;-1; 1;-1; 1; 1; 3]$ |
| $H_u^T*[-1;-1;-1;-1;-1;-1;1;1]/2=[0; 0; 0; 0; 0; 0; 0; 4]$ | $H_u^T*[-1;1;-1;-1;-1;1;1;1]/2=[-1; 1;-1; 1;-1; 1; 3; 1]$ |
| $H_u^T*[-1;1;-1;1;-1;1;-1;1]/2 =[-2; 0;-2; 0;-2; 0; 2; 0]$ | $H_u^T*[-1;1;-1;1;1;1;1;-1]/2=[-1;-1; 1; 1; 1; 1; 3;-1]$ |

Figure 12. Some numeric examples of expressions $H_u^T*\bar{e}_i/2$ where $H_u^T$ is the transposed genetic Hadamard matrix (see Figure 5) and $\bar{e}_i$ are even bi-sign vectors (in the left half of the table) and odd bi-sign vectors (in the right half of the table).

In other words here we have new complementary sets: the sets of odd vectors $\bar{e}_i$ and of odd vectors $H_u^T*\bar{e}_i/2$ are complementary each to another; the sets of even vectors $\bar{e}_i$ and of even vectors $H_u^T*\bar{e}_i/2$ are complementary each to another. One can note else that the set of even vectors $H_u^T*\bar{e}_i/2$ contains three types of coordinate magnitudes: 0, 2 and 4 (with plus or minus); and the set of odd vectors $H_u^T*\bar{e}_i/2$ contains two types of coordinate magnitudes: 1 and 3. It resembles the phenomenological situation with numbers 3 and 2 of hydrogen bonds in complementary pairs of nitrogenous bases G-C and A-U/T. Figure 13 shows an example of complementary sequences of such vectors $\bar{e}_i$ and $H_u^T*\bar{e}_i/2$ schematically.

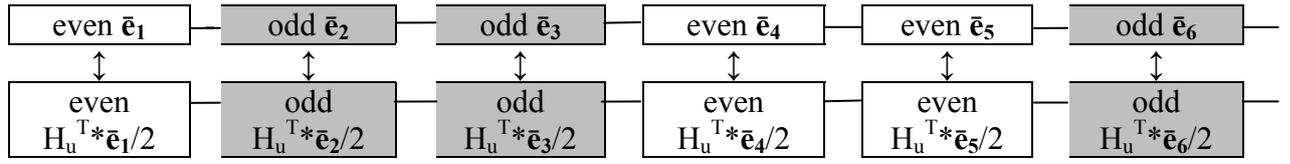

*Figure 13. A schematic example of two complementary sequences of even and odd vectors $\bar{e}_i$ and $H_u^T*\bar{e}_i/2$ where $\bar{e}_i$ are bi-sign vectors and $H_u^T$ is the transposed Hadamard matrix. Odd vectors of both types are marked by dark color.*

In the new sequence of vectors $H_u^T*\bar{e}_i/2$ one can denote each odd (even) vector by means of element "+1" ("-1") and then one can consider 8 adjacent elements of the new sequence as a new bi-sign vector $\bar{a}_i$ which belongs to a set of bi-signs vectors $\bar{a}_i$ of the second level. In this case a new sequence of even and odd bi-sign vectors $\bar{a}_i$ of the second level arises. A relevant sequence of even and odd vectors $H_u^T*\bar{a}_i/2$ of the second level can be constructed also. It is obvious that a multi-level hierarchy of such complementary sequences can be constructed by means of a similar algorithmic way if the initial sequence of vectors $\bar{e}_i$ is long enough.

These complementary sequences, which are connected with genetic Hadamard matrices, can be used to investigate some structural features of genetic sequences. It seems to be interesting because the classical complementary sequences of nitrogenous bases in DNA and RNA are very important for all genetic informatics (another example of a complementary exists in complementary pairs of histones H2A-H2B and H3-H4). Taking into account properties of Hadamard matrices and of complementary sequences described above, the author puts forward a new approach to the genetic coding which uses so called operator model (or theory) of genetic coding. The beginnings of the operator model are presented briefly below.

## 5. About the operator model of the genetic coding.

In the field of digital communication, coding of signals provides an effective and reliable transfer of information at difficult conditions of noises, etc. It is known that Hadamard matrices and Walsh functions are used in signal coding widely because they can increase a quality of information transfer. Let us consider an example of the simplest case of such coding with an application of Hadamard matrices. In this case we have an initial digital sequence $X(x_0, x_1,…x_N)$ (where N is a large enough) which should be transferred. First of all we divide this sequence X into some equal parts $a_1, a_2,..., a_k$, each of which contains $2^n$ members $x_i$ where "n" is integer number (for instance, if n=3 then the length of each part is equal to $8=2^3$ members). If the last of such parts of the sequence X has less than $2^n$ members, then a necessary quantity of zero members is added into its end. Each of these parts is interpreted as a $2^n$-dimensional vector $\bar{u}_i$. In this case the whole sequence X is represented as a sequence of such $2^n$-dimensional vectors $\bar{u}_1, \bar{u}_2,..., \bar{u}_k$. After this a Hadamard spectrum is created in a form of a new $2^n$-dimensional vector $\bar{e}_i$ (i=1, 2,…,k) for each of these $2^n$-dimensional vectors $\bar{u}_i$ by means of a Hadamard $(2^n*2^n)$-matrix H. In the result, a sequence of Hadamard spectra $\bar{e}_i$ arises which can differ drastically from the initial sequences of vectors $\bar{u}_1, \bar{u}_2,..., \bar{u}_k$. This sequence of Hadamard spectra $\bar{e}_i$ (but not the sequence of the initial signal vectors $\bar{u}_1, \bar{u}_2,..., \bar{u}_k$) is transferred from a signal transmitter to a remote receiver where the initial sequence of vectors $\bar{u}_1, \bar{u}_2,..., \bar{u}_k$ is reconstructed on the basis of the property of Hadamard matrices: $H(n)*H(n)^T = n*I_n$. The transfer of Hadamard spectra allows increasing a quality of information transfer at difficult conditions of a signal trip [Ahmed, Rao, 1975; Sklar, 2001; Trahtman, Trahtman, 1975; etc.].

In the case of genetic information, DNA can be considered as a signal transmitter, which should send out genetic messages in such form which provides a reliable and qualitative transfer of genetic information. Ribosomes (or some other participants of genetic systems), which receive nucleotide sequences to create proteins, can be considered as a remote receiver by analogy with the scheme of digital communication. The way from DNA to ribosomes is not simple and short in difficult conditions of molecular bouillon. It demands special decisions to ensure a qualitative transfer of genetic information. We suppose that the nature uses an opportunity of coding of genetic information by means of Hadamard operators and of Hadamard spectra by analogy with using of Hadamard matrices in digital communication. In other words, the living nature uses a transfer of molecular genetic information in a form of Hadamard spectra of some hidden vectors which can be considered as pro-genetic vectors. These pro-genetic vectors are numeric presentations of some "pro-genes". From this viewpoint, a pro-genetic level exists in organisms; known genetic sequences bear a set of Hadamard spectra of numeric presentations of hidden "pro-genes".

Taking all of these into account, the author puts forward the following working hypothesis:
- Genetic sequences can be presented on a basis of their molecular parameters in a form of numeric sequences of $2^n$-dimensional vectors which play in organism a role of sequences of Hadamard spectra of hidden multi-dimensional vectors. These hidden vectors can be interpreted as "pro-genetic" vectors which possess a significant physiological meaning (a title "pro-genes" is used here conditionally).

A specificity of this operator approach can be illustrated in this article briefly by an example of a genetic sequence with 21 triplets (2):

$$\text{GGC-AUC-GUU-GAA-CAG-UGU-UGC-ACA-UCU-AUC-UGC-UCU-CUU-}$$
$$\text{-UAC-CAG-CUU-GAG-AAC-UAC-UGU-AAC} \tag{2}$$

The symbolic sequence (2) can be digitalized into a sequence of elements "+1" and "-1". In other words one can consider a numeric presentation of the symbolic sequence (2) in a form of its numeric presentation (by analogy, for instance, with quantum mechanics which considers numeric presentations of states of quantum mechanical systems frequently). Let us use a presentation of each black triplet (see Figure 4) as the element "+1" and a presentation of each white triplet as the element "-1" (it is obvious that some other variants of numeric presentations of triplets are possible also but we do not consider them in this article). In this case the sequence (2) is transformed into the following bi-sign sequence (3):

$$1 \rightarrow -1 \rightarrow 1 \rightarrow -1 \rightarrow -1 \rightarrow -1 \rightarrow -1 \rightarrow 1 \rightarrow 1 \rightarrow -1 \rightarrow -1 \rightarrow 1 \rightarrow 1 \rightarrow -1 \rightarrow -1 \rightarrow 1 \rightarrow -1 \rightarrow -1 \rightarrow -1 \rightarrow -1 \rightarrow -1 \tag{3}$$

At this stage one can consider this sequence (3) with some addition as a sequence of three 8-dimensional vectors $\bar{e}_1$, $\bar{e}_2$ and $\bar{e}_3$:

$$\bar{e}_1 = [1; -1; 1; -1; -1; -1; -1; 1;]; \bar{e}_2 = [1; -1; -1; 1; 1; -1; -1; 1]; \bar{e}_3 = [-1; -1; -1; -1; -1; -1; -1; -1] \tag{4}$$

The third 8-dimensional vector $\bar{e}_3$ in (4) contains five last numbers from the numeric sequence (3) and three numbers "-1" else, which we added in its end by analogy with the method of digital communication described in the beginning of this section. The vector $\bar{e}_1$ is odd (it contains the odd number 3 of elements "+1"), and the vectors $\bar{e}_2$ and $\bar{e}_3$ are even bi-sign vectors (they contain even numbers 4 and 0 of elements "+1"). The proposed operator model interprets these bi-sign vectors $\bar{e}_1$, $\bar{e}_2$ and $\bar{e}_3$ from (4) as Hadamard spectra $H_u * \bar{u}_1/4$, $H_u * \bar{u}_2/4$, $H_u * \bar{u}_3/4$ of some hidden pro-genetic vectors $\bar{u}_1$, $\bar{u}_2$, $\bar{u}_3$:

$$\bar{e}_1 = H_u * \bar{u}_1/4; \quad \bar{e}_2 = H_u * \bar{u}_2/4 \text{ and } \bar{e}_3 = H_u * \bar{u}_3/4 \tag{5}$$

(the genetic Hadamard matrix $H_u$ is one of possible genetic Hadamard matrices which can be used in this approach). These pro-genetic vectors $\bar{u}_1$, $\bar{u}_2$ and $\bar{u}_3$ can be revealed by means of the operation $H_u^T*(\bar{e}_i)/2$:

$$\bar{u}_1 = H_u^T*(\bar{e}_1)/2 = [1; 1; -1; -1; -1; -1; -3; 1];$$
$$\bar{u}_2 = H_u^T*(\bar{e}_2)/2 = [-2; 0; -2; 0; 2; 0; -2; 0];$$
$$\bar{u}_3 = H_u^T*(\bar{e}_3)/2 = [0; -2; 0; -2; 0; -2; 0; 2] \tag{6}$$

The vector $\bar{u}_1$ is odd (it contains coordinates with odd numbers 1 and 3), and the vectors $\bar{u}_2$ and $\bar{u}_3$ are even pro-genetic vectors (it contains coordinates with even numbers 0 and 2). In this example we have one complementary pair of odd type $\bar{e}_1 \leftrightarrow \bar{u}_1$ and two complementary pairs of even type $\bar{e}_2 \leftrightarrow \bar{u}_2$, $\bar{e}_3 \leftrightarrow \bar{u}_3$ (see Figures 12 and 13).

One can think that such new complementary sequences $\bar{e}_i$ and $\bar{u}_i$ can be replicated by analogy with replications of complementary nucleotide sequences of DNA. Whether these pro-genetic vectors $\bar{u}_i$ be revealed in real molecular forms (for example, in parameters of histones) or they are virtual objects of a management level in genetic systems? This question is open now and it should be investigated in the nearest future.

The main features of the proposed operator model are the following:
- this model uses numeric presentations of symbolic genetic sequences;
- it interprets genetic sequences as sequences of $2^n$-dimensional vectors;
- it supposes an existence of "pro-genes" and of a complementary system of vectors which are connected with "pro-genes" ;
- it uses genetic matrices (first of all, genetic Hadamard matrices) which are constructed on the basis of genetic alphabets and of genetic algorithms;
- it pays a high attention to facts of complementary relations among genetic objects.

The author thinks that features of the genetic alphabets and of the degeneracy of the vertebrate mitochondria genetic code are dictated by special matrix operators which exist in the nature or which reflect structural properties of the nature.

Many authors suppose an existence of a hidden system of management of genes which participates in famous phenomena of jumping genes, splicing and many others. We think that the proposed operator model of genetic coding, some initial elements of which is described in this article, can be useful to study these phenomena. Beside of genetic Hadamard matrices, interesting applications of matrix operators exist on the base of matrices of genetic bipolar algebras (or Yin-Yang algebras [Petoukhov, 2008 a, c-e]). These operators correspond to a case when all parameters of such bipolar algebras are equal to 1 ($x_1 = x_2 = x_3 = \ldots = 1$). In our opinion the future operator theory of genetic coding should collect the whole set of matrix operators which are essential for modelling of genetic coding. The main aim of this theory is a construction of a possible bridge between molecular genetics and different fields of mathematical natural sciences where matrix operators are important. The author plans to publish more details of this operator approach to genetic coding some later.

Here one should note additionally that many variants of a possible connection between genetic sequences and Hadamard spectra exist because of the following main reasons:
- Many variants exist for a digitization of genetic sequences on a basis of real molecular parameters. For example the nitrogenous bases A, C, G, U/T possess certain quantities of hydrogen bonds; certain quantities of different atoms (carbon, nitrogen, oxygen, hydrogen); certain quantities of molecular rings (purine and pyrimidine), etc. Each type of such molecular parameters can be used to digitalize genetic sequences.
- Divisions of a genetic sequence into equal parts can be made by means of many variants also. For example, a length of such part can be equal to 2, 4, 8, 16, etc. for different variants of genetic Hadamard ($2^n*2^n$)-matrices.
- Different variants of genetic Hadamard matrices exist; some of them are interrelated each with other by means of positional and alphabetic permutations of genetic elements in genetic matrices as it was described in this article and in [Petoukhov, 2008 a-e; Petoukhov, He, 2009].

## 6. 8-dimensional hypercomplex numbers on the base of the dyadic-shift decomposition of the Rademacher form of genomatrices

The genomatrix [C A; U G]$^{(3)}$ in its numeric Rademacher form $R_R$ (Figure 5, top) can be presented in a form of sum of 8 sparse matrices $R_0, R_1, \ldots, R_7$ (Figure 14):

$$R_R = R_0 + R_1 + R_2 + R_3 + R_4 + R_5 + R_6 + R_7 =$$

$$\begin{bmatrix} 1&0&0&0&0&0&0&0\\0&1&0&0&0&0&0&0\\0&0&1&0&0&0&0&0\\0&0&0&1&0&0&0&0\\0&0&0&0&1&0&0&0\\0&0&0&0&0&1&0&0\\0&0&0&0&0&0&1&0\\0&0&0&0&0&0&0&1 \end{bmatrix} + \begin{bmatrix} 0&1&0&0&0&0&0&0\\1&0&0&0&0&0&0&0\\0&0&0&1&0&0&0&0\\0&0&1&0&0&0&0&0\\0&0&0&0&0&1&0&0\\0&0&0&0&1&0&0&0\\0&0&0&0&0&0&0&1\\0&0&0&0&0&0&1&0 \end{bmatrix} + \begin{bmatrix} 0&0&-1&0&0&0&0&0\\0&0&0&-1&0&0&0&0\\1&0&0&0&0&0&0&0\\0&1&0&0&0&0&0&0\\0&0&0&0&0&0&-1&0\\0&0&0&0&0&0&0&-1\\0&0&0&0&1&0&0&0\\0&0&0&0&0&1&0&0 \end{bmatrix} + \begin{bmatrix} 0&0&0&-1&0&0&0&0\\0&0&-1&0&0&0&0&0\\0&1&0&0&0&0&0&0\\1&0&0&0&0&0&0&0\\0&0&0&0&0&0&0&-1\\0&0&0&0&0&0&-1&0\\0&0&0&0&0&1&0&0\\0&0&0&0&1&0&0&0 \end{bmatrix} +$$

$$\begin{bmatrix} 0&0&0&0&1&0&0&0\\0&0&0&0&0&1&0&0\\0&0&0&0&0&0&-1&0\\0&0&0&0&0&0&0&-1\\1&0&0&0&0&0&0&0\\0&1&0&0&0&0&0&0\\0&0&-1&0&0&0&0&0\\0&0&0&-1&0&0&0&0 \end{bmatrix} + \begin{bmatrix} 0&0&0&0&0&1&0&0\\0&0&0&0&1&0&0&0\\0&0&0&0&0&0&0&-1\\0&0&0&0&0&0&-1&0\\0&1&0&0&0&0&0&0\\1&0&0&0&0&0&0&0\\0&0&0&-1&0&0&0&0\\0&0&-1&0&0&0&0&0 \end{bmatrix} + \begin{bmatrix} 0&0&0&0&0&0&-1&0\\0&0&0&0&0&0&0&-1\\0&0&0&-1&0&0&0&0\\0&0&0&0&-1&0&0&0\\0&0&-1&0&0&0&0&0\\0&0&0&-1&0&0&0&0\\-1&0&0&0&0&0&0&0\\0&-1&0&0&0&0&0&0 \end{bmatrix} + \begin{bmatrix} 0&0&0&0&0&0&0&-1\\0&0&0&0&0&0&-1&0\\0&0&0&0&0&-1&0&0\\0&0&0&0&-1&0&0&0\\0&0&0&-1&0&0&0&0\\0&0&-1&0&0&0&0&0\\0&-1&0&0&0&0&0&0\\-1&0&0&0&0&0&0&0 \end{bmatrix}$$

*Figure 14. The dyadic-shift decomposition of the Rademacher form (Figure 5, top) of the genomatrix [C A; U G]$^{(3)}$ into sum of 8 sparse matrices $R_0, R_1, \ldots, R_7$.*

The author revealed that this set of 8 matrices $R_0, R_1, \ldots, R_7$ (where $R_0$ is identity matrix) is closed relative to multiplication and it satisfies the table of multiplication on Figure 15.

|       | 1     | $R_1$ | $R_2$  | $R_3$  | $R_4$  | $R_5$  | $R_6$ | $R_7$ |
|-------|-------|-------|--------|--------|--------|--------|-------|-------|
| 1     | 1     | $R_1$ | $R_2$  | $R_3$  | $R_4$  | $R_5$  | $R_6$ | $R_7$ |
| $R_1$ | $R_1$ | 1     | $R_3$  | $R_2$  | $R_5$  | $R_4$  | $R_7$ | $R_6$ |
| $R_2$ | $R_2$ | $R_3$ | -1     | $-R_1$ | $-R_6$ | $-R_7$ | $R_4$ | $R_5$ |
| $R_3$ | $R_3$ | $R_2$ | $-R_1$ | -1     | $-R_7$ | $-R_6$ | $R_5$ | $R_4$ |
| $R_4$ | $R_4$ | $R_5$ | $R_6$  | $R_7$  | 1      | $R_1$  | $R_2$ | $R_3$ |
| $R_5$ | $R_5$ | $R_4$ | $R_7$  | $R_6$  | $R_1$  | 1      | $R_3$ | $R_2$ |
| $R_6$ | $R_6$ | $R_7$ | $-R_4$ | $-R_5$ | $-R_2$ | $-R_3$ | 1     | $R1$  |
| $R_7$ | $R_7$ | $R_6$ | $-R_5$ | $-R_4$ | $-R_3$ | $-R_2$ | $R_1$ | 1     |

*Figure 15. The multiplication table of basic matrices $R_0, R_1, \ldots, R_7$ (where $R_0$ is identity matrix) which corresponds to 8-dimensional algebra over the field of real numbers. It defines the 8-dimensional numeric system ("genooctetons").*

The multiplication table on Figure 15 is asymmetrical relative to the main diagonal and corresponds to non-commutative algebra of 8-dimensional hypercomplex numbers of a special type. We call these hypercomplex numbers as "dyadic-shift genetic octetons" (or briefly "R-genooctetons"; below "H-genooctetons" are described also). It should be noted that such decomposition (Figure 14) is related closely with a structure of famous (8x8)-matrices of dyadic shifts which are used widely in theory of signal processing, sequency theory by Harmut, etc. [Ahmed, Rao, 1975; Harmut, 1977, §1.2.6] and whose relations with some features of the genetic code have been described by the author previously [Petoukhov, 2008a, § 2.7; Petoukhov, 2010; Petoukhov & He, 2010, Chapter 1, Figure 5]. More precisely, in this "dyadic-shift decomposition" each basic matrix $R_k$ (k=0, 1, 2, …, 7) has its non-zero entries "+1" or "-1" only in those cells which are occupied by appropriate number "k" (from the set of numbers 0, 1, 2, 3, 4, 5, 6, 7) in the (8x8)-matrix of dyadic shifts. Multiplication sub-tables, which are shown by bold lines in the table on Figure 15, define 2-dimensional and 4-dimensional sub-algebras of these R-genooctetons. The numeric system of dyadic-shift R-genooctetons differs cardinally from the system of genetic 8-dimensional numbers which have been described by the author in matrix genetics previously [Petoukhov, 2008 a, c, d; Petoukhov & He, 2010, Section 3] and which are called "8-dimensional bipolars" or "8-dimensional Yin-Yang genonumbers".

The known term "octonions" is not appropriate for the case of the multiplication table on Figure 15 because this mathematical term is usually used for members of normed <u>division</u> algebra over the real numbers (http://en.wikipedia.org/wiki/Octonion). One should emphasize here that the 8-dimensional matrix algebra of R-genooctetons is non-division algebra because it has zero divisors. It means that such non-zero genooctetons exist whose product is equal to zero. For example ($R_3+R_4$) and ($R_2+R_5$) (see Figure 14) are non-zero R-genooctetons, but their product is equal to zero. So, R-genooctetons differ from classical octonions. Correspondingly the multiplication table of R-genooctetons (Figure 15) differs from Cayley's octonion multiplication table (http://en.wikipedia.org/wiki/Octonion).

A general form of R-genooctetons R (Figure 15) is the following:
$$R = x_0 + x_1 \ast \mathbf{R_1} + x_2 \ast \mathbf{R_2} + x_3 \ast \mathbf{R_3} + x_4 \ast \mathbf{R_4} + x_5 \ast \mathbf{R_5} + x_6 \ast \mathbf{R_6} + x_7 \ast \mathbf{R_7} \qquad (7)$$

where $x_0, x_1, \ldots, x_7$ are real numbers. Here the first component $x_0$ is a scalar. Other 7 components $x_1 \ast \mathbf{R_1}, x_2 \ast \mathbf{R_2}, x_3 \ast \mathbf{R_3}, x_4 \ast \mathbf{R_4}, x_5 \ast \mathbf{R_5}, x_6 \ast \mathbf{R_6}, x_7 \ast \mathbf{R_7}$ are non-scalar units. The multiplication table (Figure 15) shows that the product of two unlike basis units from the set $R_1, R_2, \ldots, R_7$ is equal to a third basis unit (with the sign "+" or "-"), and such triad of units forms a closed group under multiplication. In this way the following system of 7 triads appears (Figure 16, left column):

| Triads of basis elements | Modulo-2 addition of binary numbers | Decimal notations of these binary numbers |
|---|---|---|
| $R_1, R_2, R_3$ | 001+010=011 | 1+2=3 |
| $R_1, R_4, R_5$ | 001+100=101 | 1+4=5 |
| $R_1, R_6, R_7$ | 001+110=111 | 1+6=7 |
| $R_2, R_4, R_6$ | 001+100=101 | 2+4=6 |
| $R_2, R_5, R_7$ | 010+101=111 | 2+5=7 |
| $R_3, R_4, R_7$ | 011+100=111 | 3+4=7 |
| $R_3, R_5, R_6$ | 011+101=110 | 3+5=6 |

*Figure 16. Triads of basis elements of R-genooctetons (Figure 15) and the correspondence of their indexes to modulo-2 addition for binary numbers.*

This set of triads reveals a new relation of R-genooctetons with mathematical operation of modulo-2 addition which is one of fundamental operations for binary variables. Binary notations of decimal indexes 1, 2, 3, 4, 5, 6, 7 of the basis units $R_1, R_2,\ldots, R_7$ are 001, 010, 011, 100, 101, 110, 111 correspondingly. By definition, the modulo-2 addition of two numbers, which are written in binary notation, is made in bitwise manner in accordance with the following rules:

$$0 + 0 = 0,\ 0 + 1 = 1,\ 1 + 0 = 1,\ 1 + 1 = 0 \qquad (8)$$

Figure 16 (two right columns) shows that three indexes of basis units in each triad interact with each other in accordance with the rules (8) of the modulo-2 addition.

Let us return now to six possible variants of positional permutations inside triplets which were described above: 1-2-3, 2-3-1, 3-1-2, 1-3-2, 2-1-3, 3-2-1. The six mosaic matrices $P_{123}^{(3)}$, $P_{231}^{(3)}$, $P_{312}^{(3)}$, $P_{132}^{(3)}$, $P_{213}^{(3)}$, $P_{321}^{(3)}$ correspond to these variants (see details in [Petoukhov, 2008b]). Their Rademacher forms of presentation $R_{123}, R_{231}, R_{312}, R_{132}, R_{213}, R_{321}$ are shown on Figure 17.

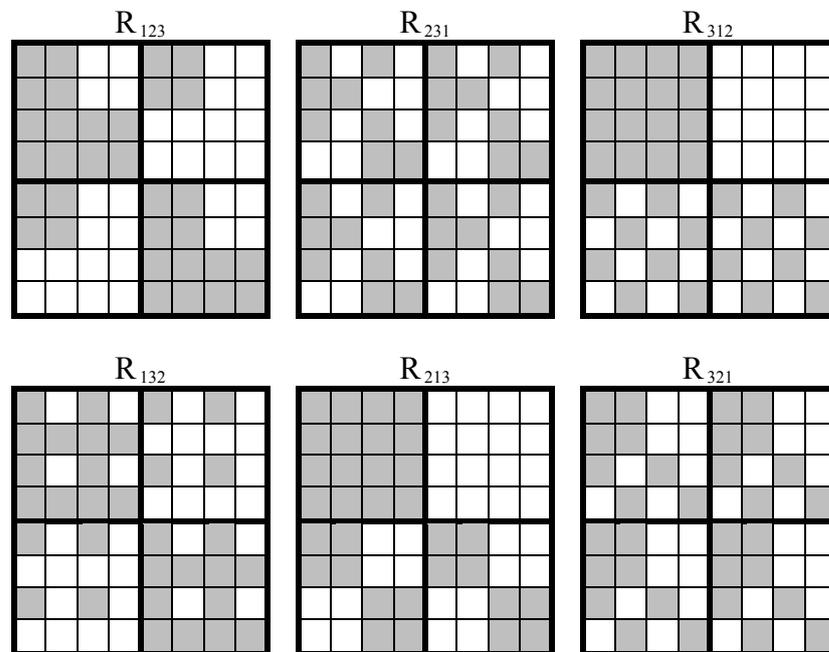

*Figure 17. Rademacher forms of genomatrices $R_{123}, R_{231}, R_{312}, R_{132}, R_{213}, R_{321}$, in which each black (white) cell means "+1" ("-1").*

One can check that the same dyadic-shift decomposition of each of these (8x8)-matrices $R_{123}$, $R_{231}, R_{312}, R_{132}, R_{213}, R_{321}$ gives six various sets of 8 basic sparse matrices $R_0, R_1, R_2, \ldots, R_7$. Each of these sets is also closed relative to multiplication. Appropriate multiplication tables for these six various sets of 8 basic matrices are identical each to other and they coincide with the table on Figure 15. It means that these six various sets of 8 basic matrices are different matrix forms of presentation of the same matrix algebra of R-genooctetons. Such situation, when matrix algebra has various matrix forms of its presentation, is known in mathematics: for example, quaternions by Hamilton have the "left" matrix form of presentation and the "right" matrix form of presentation. The basic matrices $R_i$ of the R-genooctetons are related with a known geometrical concept of reflection operators which satisfy the condition $R_i^2 = I$ (where I is identity matrix) [Vinberg, 2001, p. 224]. In view of this fact, the algebra of R-genooctetons can be interpreted to some extend as the algebra of reflection operators.

Along the way one can add about existence of the "golden" R-genoocteton:

$$G = f + f^1*R_1 + f*R_2 + f^1*R_3 + f*R_4 + f^1*R_5 + f*R_6 + f^1*R_7 \qquad (9)$$

where $f=(1+5^{0.5})/2= 1,618…$ is the irrational number of the famous golden section. This "golden" R-genoocteton has some interesting mathematical properties. For example, all coordinates of R-genooctetons $G^{2n}$ are integer numbers (for instance, $G^2= 6 + 4*R_1 + 6*R_2 + 4*R_3 + 6*R_4 + 4*R_5 + 6*R_6 + 4*R_7$), but all coordinates of genooctetons $G^{2n+1}$ are irrational numbers (here n = 1, 2, 3, …).

## 7. 8-dimensional hypercomplex numbers on the base of the dyadic-shift decomposition of Hadamard genomatrices

But what one can say about existence of matrix algebras for the analogical dyadic-shift decomposition of Hadamard genomatrices (see Figure 10)? Our investigation gives a positive answer on this question and reveals the family of 8-dimensional algebras of genetic octetons of new types in these cases. Let us demonstrate initially these results on a separate example of the Hadamard genomatrix for the case of $P^{CAUG}_{123}$ (Figure 10, left top). Figure 18 shows the appropriate dyadic-shift decomposition of this Hadamard genomatrix $H_{123}$.

| 1 | 1 | -1 | -1 | 1 | 1 | -1 | -1 |
|---|---|----|----|---|---|----|----|
| -1 | 1 | 1 | -1 | -1 | 1 | 1 | -1 |
| 1 | 1 | 1 | 1 | -1 | -1 | -1 | -1 |
| -1 | 1 | -1 | 1 | 1 | -1 | 1 | -1 |
| -1 | -1 | 1 | 1 | 1 | 1 | -1 | -1 |
| 1 | -1 | -1 | 1 | -1 | 1 | 1 | -1 |
| 1 | 1 | 1 | 1 | 1 | 1 | 1 | 1 |
| -1 | 1 | -1 | 1 | -1 | 1 | -1 | 1 |

$$G_{123} = G_0+G_1+G_2+G_3+G_4+G_5+G_6+G_7 =$$

$$\begin{pmatrix} 1 & 0 & 0 & 0 & 0 & 0 & 0 & 0 \\ 0 & 1 & 0 & 0 & 0 & 0 & 0 & 0 \\ 0 & 0 & 1 & 0 & 0 & 0 & 0 & 0 \\ 0 & 0 & 0 & 1 & 0 & 0 & 0 & 0 \\ 0 & 0 & 0 & 0 & 1 & 0 & 0 & 0 \\ 0 & 0 & 0 & 0 & 0 & 1 & 0 & 0 \\ 0 & 0 & 0 & 0 & 0 & 0 & 1 & 0 \\ 0 & 0 & 0 & 0 & 0 & 0 & 0 & 1 \end{pmatrix} + \begin{pmatrix} 0 & 1 & 0 & 0 & 0 & 0 & 0 & 0 \\ -1 & 0 & 0 & 0 & 0 & 0 & 0 & 0 \\ 0 & 0 & 0 & 1 & 0 & 0 & 0 & 0 \\ 0 & 0 & -1 & 0 & 0 & 0 & 0 & 0 \\ 0 & 0 & 0 & 0 & 0 & 1 & 0 & 0 \\ 0 & 0 & 0 & 0 & -1 & 0 & 0 & 0 \\ 0 & 0 & 0 & 0 & 0 & 0 & 0 & 1 \\ 0 & 0 & 0 & 0 & 0 & 0 & -1 & 0 \end{pmatrix} + \begin{pmatrix} 0 & 0 & -1 & 0 & 0 & 0 & 0 & 0 \\ 0 & 0 & 0 & -1 & 0 & 0 & 0 & 0 \\ 1 & 0 & 0 & 0 & 0 & 0 & 0 & 0 \\ 0 & 1 & 0 & 0 & 0 & 0 & 0 & 0 \\ 0 & 0 & 0 & 0 & 0 & 0 & -1 & 0 \\ 0 & 0 & 0 & 0 & 0 & 0 & 0 & -1 \\ 0 & 0 & 0 & 0 & 1 & 0 & 0 & 0 \\ 0 & 0 & 0 & 0 & 0 & 1 & 0 & 0 \end{pmatrix} + \begin{pmatrix} 0 & 0 & 0 & -1 & 0 & 0 & 0 & 0 \\ 0 & 0 & 1 & 0 & 0 & 0 & 0 & 0 \\ 0 & 1 & 0 & 0 & 0 & 0 & 0 & 0 \\ -1 & 0 & 0 & 0 & 0 & 0 & 0 & 0 \\ 0 & 0 & 0 & 0 & 0 & 0 & 0 & -1 \\ 0 & 0 & 0 & 0 & 0 & 0 & 1 & 0 \\ 0 & 0 & 0 & 0 & 0 & 1 & 0 & 0 \\ 0 & 0 & 0 & 0 & -1 & 0 & 0 & 0 \end{pmatrix} +$$

$$\begin{pmatrix} 0 & 0 & 0 & 0 & 1 & 0 & 0 & 0 \\ 0 & 0 & 0 & 0 & 0 & 1 & 0 & 0 \\ 0 & 0 & 0 & 0 & 0 & 0 & -1 & 0 \\ 0 & 0 & 0 & 0 & 0 & 0 & 0 & -1 \\ -1 & 0 & 0 & 0 & 0 & 0 & 0 & 0 \\ 0 & -1 & 0 & 0 & 0 & 0 & 0 & 0 \\ 0 & 0 & 1 & 0 & 0 & 0 & 0 & 0 \\ 0 & 0 & 1 & 0 & 0 & 0 & 0 & 0 \end{pmatrix} + \begin{pmatrix} 0 & 0 & 0 & 0 & 0 & 1 & 0 & 0 \\ 0 & 0 & 0 & -1 & 0 & 0 & 0 & 0 \\ 0 & 0 & 0 & 0 & 0 & 0 & 0 & -1 \\ 0 & 0 & 0 & 0 & 0 & 0 & 1 & 0 \\ 0 & -1 & 0 & 0 & 0 & 0 & 0 & 0 \\ 1 & 0 & 0 & 0 & 0 & 0 & 0 & 0 \\ 0 & 0 & 0 & 1 & 0 & 0 & 0 & 0 \\ 0 & 0 & -1 & 0 & 0 & 0 & 0 & 0 \end{pmatrix} + \begin{pmatrix} 0 & 0 & 0 & 0 & 0 & 0 & -1 & 0 \\ 0 & 0 & 0 & 0 & 0 & 0 & 0 & -1 \\ 0 & 0 & 0 & -1 & 0 & 0 & 0 & 0 \\ 0 & 0 & 0 & 0 & -1 & 0 & 0 & 0 \\ 0 & 0 & 1 & 0 & 0 & 0 & 0 & 0 \\ 0 & 0 & 0 & 1 & 0 & 0 & 0 & 0 \\ 1 & 0 & 0 & 0 & 0 & 0 & 0 & 0 \\ 0 & 1 & 0 & 0 & 0 & 0 & 0 & 0 \end{pmatrix} + \begin{pmatrix} 0 & 0 & 0 & 0 & 0 & 0 & 0 & -1 \\ 0 & 0 & 0 & 0 & 0 & 0 & 1 & 0 \\ 0 & 0 & 0 & 0 & 0 & -1 & 0 & 0 \\ 0 & 0 & 0 & 0 & 1 & 0 & 0 & 0 \\ 0 & 0 & 0 & 1 & 0 & 0 & 0 & 0 \\ 0 & 0 & -1 & 0 & 0 & 0 & 0 & 0 \\ 0 & 1 & 0 & 0 & 0 & 0 & 0 & 0 \\ -1 & 0 & 0 & 0 & 0 & 0 & 0 & 0 \end{pmatrix}$$

*Figure 18. The Hadamard genomatrix $H_{123}$ (from Figure 10, left top) and its dyadic-shift decomposition into sum of 8 sparse matrices $G_0, G_1,…, G_7$.*

This set of 8 sparse matrices $G_0, G_1,…, G_7$ (where $G_0$ is identity matrix) is closed relative to multiplication and it satisfies the multiplication table on Figure 19.

|   | 1   | $G_1$ | $G_2$ | $G_3$ | $G_4$ | $G_5$ | $G_6$ | $G_7$ |
|---|-----|-------|-------|-------|-------|-------|-------|-------|
| 1 | 1   | $G_1$ | $G_2$ | $G_3$ | $G_4$ | $G_5$ | $G_6$ | $G_7$ |
| $G_1$ | $G_1$ | -1 | $G_3$ | $-G_2$ | $G_5$ | $-G_4$ | $G_7$ | $-G_6$ |
| $G_2$ | $G_2$ | $G_3$ | -1 | $-G_1$ | $-G_6$ | $-G_7$ | $G_4$ | $G_5$ |
| $G_3$ | $G_3$ | $-G_2$ | $-G_1$ | 1 | $-G_7$ | $G_6$ | $G_5$ | $-G_4$ |
| $G_4$ | $G_4$ | $G_5$ | $G_6$ | $G_7$ | -1 | $-G_1$ | $-G_2$ | $-G_3$ |
| $G_5$ | $G_5$ | $-G_4$ | $G_7$ | $-G_6$ | $-G_1$ | 1 | $-G_3$ | $G_2$ |
| $G_6$ | $G_6$ | $G_7$ | $-G_4$ | $-G_5$ | $G_2$ | $G_3$ | -1 | $-G_1$ |
| $G_7$ | $G_7$ | $-G_6$ | $-G_5$ | $G_4$ | $G3$ | $-G_2$ | $-G_1$ | 1 |

*Figure 19. The multiplication table of basic matrices $G_0, G_1,… , G_7$ (where $G_0$ is identity matrix) for Hadamard genomatrix $H_{123}$ (see Figure 18)*

The asymmetrical multiplication table (Figure 19) corresponds to a special type of 8-dimensional hypercomplex numbers. We call them conditionally as "H-genooctetons". Their algebra is non-commutative and non-division algebra (it has zero divisors, for example, $G_3+G_4$ is non-zero matrix but $(G_3+G_4)^2$ is equal to zero). The Hadamard genomatrix $H_{123}$ itself is one of 8-dimensional H-genooctetons. The multiplication table (Figure 19) shows that the product of two unlike basis units from the set $G_1, G_2, … , G_7$ is equal to a third unit from this set (with the sign "+" or "-"), and such triad of units forms a closed group under multiplication. In this way the following system of 7 triads appears again (by analogy with Figure 16): $G_1,G_2,G_3$; $G_1,G_4,G_5$; $G_1,G_6,G_7$; $G_2,G_4,G_6$; $G_2,G_5,G_7$; $G_3,G_4,G_7$; $G_3,G_5,G_6$. The multiplication table (Figure 19) has 28 signs "-" and 36 signs "+"; the disposition of these signs is identical to the disposition of signs in one of Hadamard matrices but the dyadic-shift decomposition of the relevant Hadamard matrix does not correspond to any system of 8-dimensional hypercomplex numbers.

It should be emphasized that only some of a huge quantity of Hadamard (8x8)-matrices in their dyadic-shift decompositions give the basis for matrix forms of presentation of 8-dimensional hypercomplex numbers. In other words, the most of Hadamard (8x8)-matrices in their dyadic-shift decomposition have no any relation with 8-dimensional hypercomplex numbers. For example, the dyadic-shift decompositions of symmetrical Hadamard (8*8)-matrices (Figure 2), which are used in technical digital communication frequently, have no any relation with 8-dimensional hypercomplex numbers. In contrast to modern techniques of digital communication, the genetic system is related with such Hadamard (8*8)-matrices, whose dyadic-shift decompositions are associated with 8-dimensional hypercomplex numbers (genooctetons).

But what one can say about the case of the transposed Hadamard genomatrix $H_{123}^T$? We study this case with the following result: the dyadic-shift decomposition of the transposed Hadamard genomatrix $H_{123}^T$ has its own matrix algebra of 8-dimensional hypercomplex numbers whose multiplication table is the transposed analogue of the multiplication table (Figure 19) for the case of the Hadamard genomatrix $H_{123}$.

Now let us take another Hadamard genomatrix which is generated from the Hadamard genomatrix $H_{123}$ by means of positional permutations inside triplets, for example, $H_{231}$ (Figure 9 and Figure 10, top right). Its dyadic-shift decomposition gives a new set of 8 sparse matrices $B_0$, $B_1,.. , B_7$ (Figure 20).

$$H_{231} = B_0+B_1+B_2+B_3+B_4+B_5+B_6+B_7 =$$

$$\begin{bmatrix} 1 & 0 & 0 & 0 & 0 & 0 & 0 & 0 \\ 0 & 1 & 0 & 0 & 0 & 0 & 0 & 0 \\ 0 & 0 & 1 & 0 & 0 & 0 & 0 & 0 \\ 0 & 0 & 0 & 1 & 0 & 0 & 0 & 0 \\ 0 & 0 & 0 & 0 & 1 & 0 & 0 & 0 \\ 0 & 0 & 0 & 0 & 0 & 1 & 0 & 0 \\ 0 & 0 & 0 & 0 & 0 & 0 & 1 & 0 \\ 0 & 0 & 0 & 0 & 0 & 0 & 0 & 1 \end{bmatrix} + \begin{bmatrix} 0 & -1 & 0 & 0 & 0 & 0 & 0 & 0 \\ 1 & 0 & 0 & 0 & 0 & 0 & 0 & 0 \\ 0 & 0 & 0 & -1 & 0 & 0 & 0 & 0 \\ 0 & 0 & 1 & 0 & 0 & 0 & 0 & 0 \\ 0 & 0 & 0 & 0 & 0 & -1 & 0 & 0 \\ 0 & 0 & 0 & 0 & 1 & 0 & 0 & 0 \\ 0 & 0 & 0 & 0 & 0 & 0 & 0 & -1 \\ 0 & 0 & 0 & 0 & 0 & 0 & 1 & 0 \end{bmatrix} + \begin{bmatrix} 0 & 0 & 1 & 0 & 0 & 0 & 0 & 0 \\ 0 & 0 & 0 & -1 & 0 & 0 & 0 & 0 \\ -1 & 0 & 0 & 0 & 0 & 0 & 0 & 0 \\ 0 & 1 & 0 & 0 & 0 & 0 & 0 & 0 \\ 0 & 0 & 0 & 0 & 0 & 0 & 1 & 0 \\ 0 & 0 & 0 & 0 & 0 & 0 & 0 & -1 \\ 0 & 0 & 0 & 0 & -1 & 0 & 0 & 0 \\ 0 & 0 & 0 & 0 & 0 & 1 & 0 & 0 \end{bmatrix} + \begin{bmatrix} 0 & 0 & 0 & -1 & 0 & 0 & 0 & 0 \\ 0 & 0 & -1 & 0 & 0 & 0 & 0 & 0 \\ 0 & 1 & 0 & 0 & 0 & 0 & 0 & 0 \\ 1 & 0 & 0 & 0 & 0 & 0 & 0 & 0 \\ 0 & 0 & 0 & 0 & 0 & 0 & 0 & -1 \\ 0 & 0 & 0 & 0 & 0 & 0 & -1 & 0 \\ 0 & 0 & 0 & 0 & 0 & 1 & 0 & 0 \\ 0 & 0 & 0 & 0 & 1 & 0 & 0 & 0 \end{bmatrix} +$$

$$\begin{bmatrix} 0 & 0 & 0 & 0 & 1 & 0 & 0 & 0 \\ 0 & 0 & 0 & 0 & 0 & 1 & 0 & 0 \\ 0 & 0 & 0 & 0 & 0 & 0 & 1 & 0 \\ 0 & 0 & 0 & 0 & 0 & 0 & 0 & 1 \\ -1 & 0 & 0 & 0 & 0 & 0 & 0 & 0 \\ 0 & -1 & 0 & 0 & 0 & 0 & 0 & 0 \\ 0 & 0 & -1 & 0 & 0 & 0 & 0 & 0 \\ 0 & 0 & 0 & -1 & 0 & 0 & 0 & 0 \end{bmatrix} + \begin{bmatrix} 0 & 0 & 0 & 0 & 0 & -1 & 0 & 0 \\ 0 & 0 & 0 & 0 & 1 & 0 & 0 & 0 \\ 0 & 0 & 0 & 0 & 0 & 0 & 0 & -1 \\ 0 & 0 & 0 & 0 & 0 & 0 & 1 & 0 \\ 0 & 1 & 0 & 0 & 0 & 0 & 0 & 0 \\ -1 & 0 & 0 & 0 & 0 & 0 & 0 & 0 \\ 0 & 0 & 0 & 1 & 0 & 0 & 0 & 0 \\ 0 & 0 & -1 & 0 & 0 & 0 & 0 & 0 \end{bmatrix} + \begin{bmatrix} 0 & 0 & 0 & 0 & 0 & 0 & 1 & 0 \\ 0 & 0 & 0 & 0 & 0 & 0 & 0 & -1 \\ 0 & 0 & 0 & 0 & -1 & 0 & 0 & 0 \\ 0 & 0 & 0 & 0 & 0 & 1 & 0 & 0 \\ 0 & 0 & -1 & 0 & 0 & 0 & 0 & 0 \\ 0 & 0 & 0 & 1 & 0 & 0 & 0 & 0 \\ 1 & 0 & 0 & 0 & 0 & 0 & 0 & 0 \\ 0 & -1 & 0 & 0 & 0 & 0 & 0 & 0 \end{bmatrix} + \begin{bmatrix} 0 & 0 & 0 & 0 & 0 & 0 & 0 & -1 \\ 0 & 0 & 0 & 0 & 0 & 0 & -1 & 0 \\ 0 & 0 & 0 & 0 & 0 & 1 & 0 & 0 \\ 0 & 0 & 0 & 0 & 1 & 0 & 0 & 0 \\ 0 & 0 & 0 & 1 & 0 & 0 & 0 & 0 \\ 0 & 0 & 1 & 0 & 0 & 0 & 0 & 0 \\ 0 & -1 & 0 & 0 & 0 & 0 & 0 & 0 \\ -1 & 0 & 0 & 0 & 0 & 0 & 0 & 0 \end{bmatrix}$$

*Figure 20. The dyadic-shift decomposition of the Hadamard genomatrix $H_{231}$ (Figure 9 and Figure 10, top right).*

This set is also closed relative to multiplication and it satisfies the new multiplication table (Figure 21).

|  | 1 | $B_1$ | $B_2$ | $B_3$ | $B_4$ | $B_5$ | $B_6$ | $B_7$ |
|---|---|---|---|---|---|---|---|---|
| 1 | 1 | $B_1$ | $B_2$ | $B_3$ | $B_4$ | $B_5$ | $B_6$ | $B_7$ |
| $B_1$ | $B_1$ | -1 | $-B_3$ | $B_2$ | $B_5$ | $-B_4$ | $-B_7$ | $B_6$ |
| $B_2$ | $B_2$ | $B_3$ | -1 | $-B_1$ | $B_6$ | $B_7$ | $-B_4$ | $-B_5$ |
| $B_3$ | $B_3$ | $-B_2$ | $B_1$ | -1 | $B_7$ | $-B_6$ | $B_5$ | $-B_4$ |
| B4 | $B_4$ | $B_5$ | $B_6$ | $B_7$ | -1 | $-B_1$ | $-B_2$ | $-B_3$ |
| $B_5$ | $B_5$ | $-B_4$ | $-B_7$ | $B_6$ | $-B_1$ | 1 | $B_3$ | $-B_2$ |
| $B_6$ | $B_6$ | $B_7$ | $-B_4$ | $-B_5$ | $-B_2$ | $-B_3$ | 1 | $B_1$ |
| $B_7$ | $B_7$ | $-B_6$ | $B_5$ | $-B_4$ | $-B_3$ | $B_2$ | $-B_1$ | 1 |

*Figure 21. The multiplication table of basis matrices $B_0, B_1, …, B_7$ (where $B_0$ is identity matrix) for Hadamard genomatrix $H_{231}$ (Figure 9 and Figure 10, top right).*

More detailed data of investigations of the family of genetic 8-dimensional hypercomplex numbers (on the basis of dyadic-shift decompositions of genetic matrices) will be published later together with some ideas about possible biological, informational and physical meanings of these phenomenological facts.

## 8. Discussion

The phenomenological facts described above reveal a close connection of the genetic code (including, firstly, the phenomenon of the degeneracy of the vertebrate mitochondrial genetic code and, secondly, described permutation properties of the genetic code) with a special set of Hadamard matrices. This connection between Hadamard matrices and the genetic code is interesting in a few aspects. Genetic molecules are objects of quantum mechanics, where normalized Hadamard matrices play important role as unitary operators (it is known that an evolution of closed quantum system is described by unitary transformation). In particular, quantum computers use these matrices in a role of Hadamard gates [Nielsen, Chuang, 2001]. In view of this, some new theoretical possibilities are revealed to transfer achievements of quantum computer conceptions into the field of molecular genetics and to consider genetic system as a quantum computer. From the viewpoint of quantum mechanics and its unitary operators, first of all, Hadamard operators, a possible answer on the fundamental question about reasons for the nature to choose the four-letters genetic alphabet is the following one: the possible reason is that Hadamard (2*2)-matrices, which are the simplest unitary matrices in two-dimensional space, consist of four elements. It seems very probably that principles of quantum mechanics and quantum computers underlie structural peculiarities of the genetic code.

One can suppose that Hadamard genomatrices can be used in genetic systems by analogy with applications of Hadamard matrices in different fields of science and technology: signal processing, error-correcting and other codes, spectral analysis, multi-channel spectrometers, etc. Rows of Hadamard matrices represent orthogonal systems of Walsh functions. Such orthogonal system can be a natural base to organize storage and transfer of genetic information with noise-immunity properties by means of spectral presentations of genetic sequences on the base of these orthogonal systems and by means of using of proper codes (orthogonal, bi-orthogonal, etc.). In particular, advantages of Hadamard matrices can be exploited in genetic system for spectral analysis of genetic sequences and for utilizing emission spectra of genetic elements. It should be emphasized that a few authors have revealed already orthogonal systems of Walsh functions in macro-physiological systems, which should be agreed structurally with genetic system for their transferring along a chain of generations [Shiozaki, 1980; Carl, 1974; Ginsburg et all, 1974].

It seems that the proposed operator approach to genetic coding can be used in different branches of bioinformatics later. It is known that biological organisms possess multi-channel schemes of information transfer to provide their poly-functional activities (for example, a human organism possesses visual, audio, tactile and some other sensory channels which are inherited and which are interrelated to provide its physiological activity). A wide set of described variants of the connection of genetic sequences with Hadamard spectra can be used in organism for its multi-channel transfer of information and for multi-channel control of genetic information to provide inherited schemes of multi-channel activities of the whole organism.

Concerning the theme of 8-dimensional genetic octetons in the end of this article, one should mention that systems of 8-dimensional hypercomplex numbers (first of all, octonions and split-octonions) are one of key objects of mathematical natural sciences today. They relate to a number of exceptional structures in mathematics, among them the exceptional Lie groups; they have applications in many fields such as string theory, special relativity, the supersymmetric quantum mechanics, quantum logic, etc. (see for example, http://en.wikipedia.org/wiki/Octonion; http://en.wikipedia.org/wiki/Split-octonion; Dixon, 1994; Dray & Manoque, 2009; Lisi, 2007; Nurowski, 2009). The term "octet" is also used frequently in phenomenologic laws of modern science: the Eightfold way by M.Gell-Mann and Y.Ne'eman in physics; the octet rule in chemistry (http://en.wikipedia.org/wiki/Octet_rule), etc. In view of these facts one can think that R-genooctetons and H-genooctetons will be one of interesting parts of mathematical natural sciences in the nearest future.

Rows of Hadamard matrices correspond to orthogonal systems of Walsh functions which play the main role in fruitful sequency theory by Harmut in signal processing [Harmut, 1977, 1989]. Rows of Hadamard genomatrices correspond to very special kinds of Walsh functions ("genosystems" of Walsh functions) which define special ("genetic") variants of sequency analysis in signal processing. The author believes that this "genetic" sequency analysis, whose bases have been revealed in the field of matrix genetics, can be a key to understand important features not only of genetic informatics but also of many other inherited physiological systems (informational, morphological, etc.). In comparison with spectral analysis by means of sine waves, which is applicable to linear time-invariant systems, the sequency analysis is based on non-sinusoidal waves and is used to study systems which are changed in time (biological systems belong to such systems) [Harmut, 1977, 1989]. The author believes that inherited mechanisms of biological morphogenesis are closely associated with spatial and temporal filters from the field of geno-sequency analysis (the general theory of spatial and temporary filters of the sequency theory has been described by Harmut). The theory of morphogenetic field can be developed on the basis of these formalisms and phenomenologic facts. From this viewpoint, many morphogenetic abnormalities, which are inherited or are induced by external influences, are consequences of disruptions in physiological spatial and temporal filters of geno-sequency types. Taking into account the sequency theory by Harmut together with our data about Hadamard genomatrices and genooctetons, one can assume that biological evolution can be interpreted largely like the evolution of spatial and temporal filters of geno-sequency types. In this direction of thoughts, the author develops new approaches in bioinformatics, bioengineering and medicine.

**Acknowledgments**: Described researches were made by the author in the frame of a long-term cooperation between Russian and Hungarian Academies of Sciences and in the frames of programs of "International Society of Symmetry in Bioinformatics" (USA, http://polaris.nova.edu/MST/ISSB) and of "International Symmetry Association" (Hungary, http://symmetry.hu/). The author is grateful to Frolov K.V., Darvas G., Kappraff J., Ne'eman Y., He M., Smolianinov V.V., Vladimirov Y.S., Adamson G., Katanov D., Cristea P. for their support.